\documentclass[aps,prl,twocolumn,showpacs]{revtex4}
\usepackage{amssymb}

\usepackage{graphicx}

\begin{document}

\title{Anomalous Properties in the Normal and Superconducting States of LaRu$_3$Si$_2$}

\author{Sheng Li$^1$, Bin Zeng$^2$, Xiangang Wan$^1$, Fei Han$^2$, Jian Tao$^1$,  Huan Yang$^1$, Zhihe Wang$^1$, and Hai-Hu Wen$^1$}\email{hhwen@nju.edu.cn}

\affiliation{$^1$ Center for Superconducting Physics and Materials, National Laboratory of Solid State Microstructures
and Department of Physics, Nanjing University, Nanjing 210093,
China}

\affiliation{$^2$ Institute of Physics and Beijing National
Laboratory for Condensed Matter Physics, Chinese Academy of
Sciences, P.O. Box 603, Beijing 100190, China}

\begin{abstract}
Superconductivity in LaRu$_3$Si$_2$ with the honeycomb structure
of Ru atoms has been investigated. It is found that the normal
state specific heat C/T exhibits a deviation from the Debye model
down to the lowest temperature. A relation $C/T = \gamma_n+\beta
T^2-ATlnT$ which concerns the electron correlations can fit the
data very well. The suppression to the superconductivity by the
magnetic field is not the mean-field like, which is associated
well with the observation of strong superconducting fluctuations.
The field dependence of the induced quasiparticle density of
states measured by the low temperature specific heat shows a
non-linear feature, indicating the significant contributions given
by the delocalized quasiparticles.
\end{abstract}

\pacs{74.70.Dd, 74.25.Bt, 74.25.F-,74.10.+v} \maketitle

Superconductivity arising from non-phonon mediated pairing, such
as through exchanging the magnetic spin fluctuations, has renewed
interests in condensed matter physics. The superconducting (SC)
mechanism of the cuprates\cite{HTSMuller} and the iron
pnictides\cite{Hosono}, although not yet settled completely,
should have a close relationship with the electron
correlations.\cite{PWAnderson,Scalapino,Canfield1} A similar
assessment may extend to many others, like heavy
Fermion\cite{SiQM1} and organic materials\cite{Dressel}. In this
regard, the systems $RT_3$Si$_2$ or $RT_3$B$_2$ ($R$ stands for
the rare earth elements, like La, Ce, Y, etc., $T$ for the
transition metals, like Ru, Co and Ni, etc.) provide an
interesting platform, since a variety of combinations of chemical
compositions allow the system to be tuned between superconducting
and magnetic, and sometimes both phases
coexist.\cite{Escorne,KuHQ} Among these samples, the
LaRu$_3$Si$_2$ has a SC transition temperature as high as 7.8
K.\cite{Barz} The material of LaRu$_3$Si$_2$ contains layers of Ru
with the honeycomb structure sandwiched by the layers of La and
Si, forming a $P$6$_3$/m or $P$6$_3$ space group. Preliminary
experiment found that the SC transition temperature drops only 1.4
K with the substitution of 16 $\%$ La by Tm (possessing a magnetic
moment of about 8$\mu_B$), suggesting that the superconductivity
is robust against the local paramagnetic moment\cite{Escorne}. By
doping the La with Gd, a coexistence of superconductivity and the
spin glass state\cite{Godart} was observed. In CeRu$_3$Si$_2$, the
SC transition temperature drops to about 1 K and a valence
fluctuation model was proposed for the
pairing\cite{Rauchschwalbe}. Since the Ru atom locates just below
the Fe in the periodic table, a key player in the iron pnictide
superconductors, therefore it is very curious to know whether the
superconductivity here is induced by the electron-phonon coupling,
or by other novel mechanism, such as the electron correlations. In
this paper we report the results of transport and specific heat on
samples of LaRu$_3$Si$_2$. Our results reveal some novelties in
both the SC and normal states of LaRu$_3$Si$_2$.

\begin{figure}
\includegraphics[width=8cm]{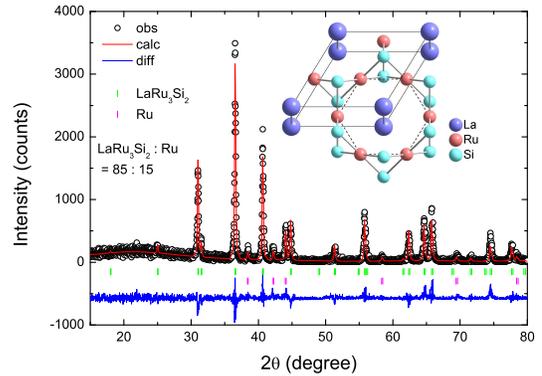}
\caption {(color online) X-ray diffraction patterns of the sample
LaRu$_3$Si$_2$. All main diffraction peaks can be indexed well by
a hexagonal structure with a = 5.68 $\AA$ and c = 7.13 $\AA$ with
Ru as the impurity phase. For some peaks the difference between
the data and the fitting is a bit large because some of the grains
of the polycrystalline sample are slightly oriented. The ratio
between LaRu$_3$Si$_2$ and Ru is found to be 85:15. The inset gives a sketch of the structure. One unit cell is highlighted by the rhomboic block.} \label{fig1}
\end{figure}

\begin{figure}
\includegraphics[width=8cm]{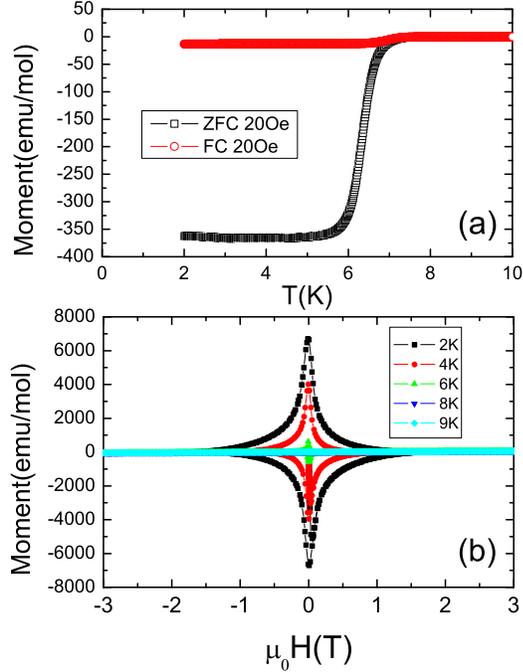}
\caption {(color online) (a) Temperature dependence of the DC
magnetization measured in the ZFC mode and the
FC mode at a magnetic field of 20 Oe. (b) The
MHLs measured with a field sweeping rate
of 50 Oe/s at different temperatures. At 9 K, the MHL shows a rough linear paramagnetic behavior.} \label{fig2}
\end{figure}

\begin{figure}
\includegraphics[width=7cm]{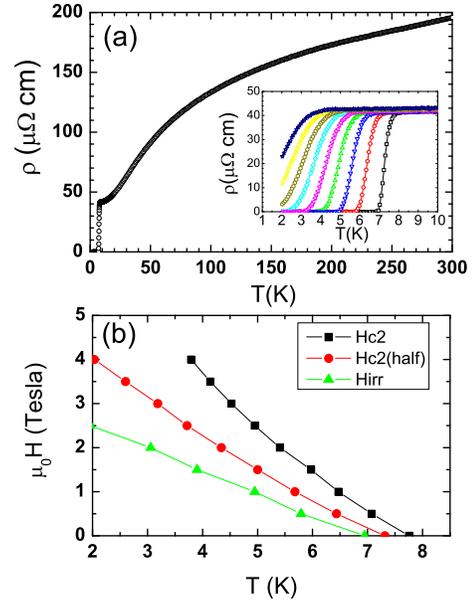}
\caption {(color online) Temperature dependence of resistivity at
zero magnetic field. The inset shows the resistivity at different
magnetic fields: 0, 0.5, 1.0, 1.5, 2.0, 2.5, 3.0, 3.5, and 4.0 T.
(b) Temperature dependence of the critical magnetic field with
three different criterions: H$_{c2}$ (squares, 95$\% \rho_n$),
H$_{c2}$ (circles, 50$\% \rho_n$) and the irreversibility line
H$_{irr}$ (up-triangles, 0.1$\% \rho_n$). There is a large area
between the H$_{c2}$ (95$\% \rho_n$) and H$_{irr}$ (0.1$\%
\rho_n$), which is probably induced by the strong SC fluctuation.}
\label{fig3}
\end{figure}

The samples were fabricated by the arc melting
method.\cite{Escorne,Barz,Godart} The starting materials La metal
pieces (99$\%$, Alfa Aesar), Ru powder (99.9$\%$) and Si powder
(99.99$\%$) were weighed and mixed well, and presseed into a
pellet in a glover box filled with Ar atmosphere (water and the
oxygen compositions were below 0.5 PPM). In order to avoid the
formation of the LaRu$_2$Si$_2$ phase, we intentionally let a
small amount of extra Ru with the nominal compositions as
LaRu$_{3+x}$Si$_2$. Three round of welding with the alternative
upper and bottom on the pellet was taken in order to achieve the
uniformity. After these refined processes, the resultant sample
contains mainly the phase of LaRu$_{3}$Si$_2$ and small amount of
Ru remains as the impurity phase. In Fig.1 we plot the x-ray
diffraction patterns (XRD) on one typical sample and the Rietveld
fitting using the GSAS program\cite{GSAS}. It is clear that the
main diffraction peaks can be indexed well by a hexagonal
structure with a = 5.68 $\AA$ and c = 7.13 $\AA$. Some weak peaks
arising from  the impurity phase Ru can also be seen. A detailed
fitting to the structural data find that the ratio between
LaRu$_3$Si$_2$ and Ru is around 85:15 for this typical sample. The
sample preparation and the quality characterized by the SC
transitions can be repeated quite well. It is found that, some of
the LaRu$_2$Si$_2$ phase with a tetragonal structure can be found
if the starting material has the nominal composition of
LaRu$_3$Si$_2$. In this case, the XRD data exhibit clearly two set
of structures and can be easily indexed by the GSAS program. For
the present sample, the absolute difference between the
experimental data and the fitting curve can be observed for some
peaks because part of the grains in the sample are slightly
aligned. The resistivity was measured with a Quantum Design
instrument PPMS-16T with a standard four-probe technique, while
the magnetization was detected by the Quantum Design instrument
SQUID-VSM with a resolution of about 5 $\times$ 10$^{-8}$ emu.

In Fig. 2(a) we present the temperature dependence of
magnetization measured in the zero-field-cooling mode (ZFC) and
the field-cooling mode (FC). By considering the demagnetization
factor on the ZFC data, the Meissner screening is estimated to be
almost 100 $\%$. This indicates that the SC connections between
the grains of LaRu$_3$Si$_2$ are very good, although we have
slight secondary phase of Ru. The onset T$_c$ determined from the
magnetization is around 7.8 K. The majority of the SC transition
occurs at about 6.6 K under a magnetic field of 20 Oe. This
difference is not induced by the inhomogeneity of the sample, it
may be induced by the relatively strong SC fluctuations (see
below). Fig.2(b) shows the magnetization hysteresis loops (MHL)
measured at different temperatures. The symmetric and clear
opening of the MHLs indicate that it is a type-II superconductor.
A roughly linear MHL was observed at 9 K, just above T$_c$,
indicating that the normal state has no long range ferromagnetic
order. We didn't observe a magnetization enhancement near T$_c$,
which was reported in Tm and Gd doped samples in early
publications\cite{Escorne,Godart}. Fig.3(a) shows the resistive
transitions at zero field (main panel) and different magnetic
fields under 4 Tesla (inset). The onset resistive transition
temperature is at 7.9 K (95$\%$ normal state resistivity
$\rho_n$), and the zero resistivity was achieved at about 6.8 K.
By applying a magnetic field, the resistive transition broadens.
Taking different criterions of resistivity we determined the upper
critical field H$_{c2} (95\% \rho_n)$, H$_{c2} (50\% \rho_n)$, and
the irreversibility line H$_{irr} (0.1\% \rho_n)$. It is clear
that there is a large difference between the H$_{c2} (95\%
\rho_n)$ and H$_{irr}$(T). We will argue that this may be induced
by the strong SC fluctuations.
\begin{figure}
\includegraphics[width=9cm]{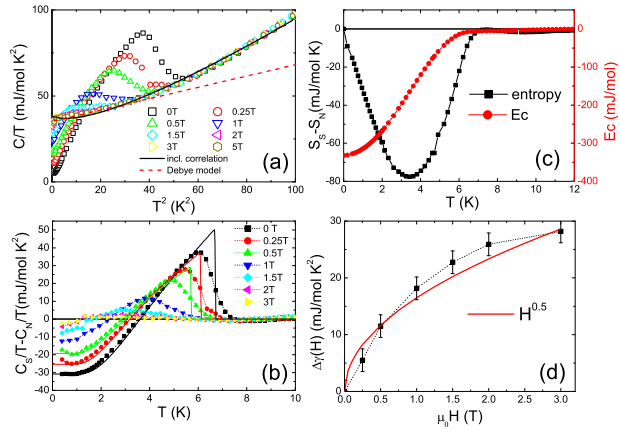}
\caption {(color online) (a) The raw data of specific heat
coefficient C/T vs. T$^2$, at different magnetic fields ranging from
0 to 5 T. The normal state data (at 5 T) shows a non-linear feature
down to the lowest temperature here, indicating a deviation from the
Debye model, as shown by the red dashed line. The solid line
represents the fit to the formula including the electron
correlations (see text). (b) The electronic specific heat
coefficient obtained by subtracting normal state value C$_N$/T (data
at 5 T) from the total. The solid lines are the theoretical fitting
curves based on the BCS model. (c) The entropy difference (squares)
between the SC state and the normal state, derived from $S_S-S_N$ =
$\int _0^T (C_S/T'-C_N/T')dT'$ at zero field and 5 T. The
condensation energy is calculated by $E_c=\int _0^T(S_S-S_N)dT'$.
(d) The magnetic field dependence of the field induced electronic
specific heat $\Delta \gamma$(H). The non-linear field dependence is
very clear. The red solid line is a fit to the $\sqrt{H}$. }
\label{fig4}
\end{figure}

The raw data of specific heat was shown in Fig.4(a). A SC anomaly
appears at about 7.6 K. Since the Ru has a $T_c$ at 0.49 K and a
quite small normal state specific heat coefficient ($\gamma_n^{Ru}$
= $2.8 mJ/mol K^2$), a slight correction of about 0.42 $mJ/mol K^2$
was made to the data. By applying a magnetic field, the SC anomaly
shifts to lower temperatures. It is interesting to note that the
transition is not shifted parallel down to the low temperatures (the
so-called mean-field like), rather the SC anomaly is suppressed.
This kind of suppression was clearly seen in the cuprate
superconductors Pr$_{0.88}$LaCe$_{0.12}$CuO$_{4-\delta}$\cite{PNAS}
and was ascribed to a strong SC fluctuation. Combining with the
resistive broadening under a magnetic field, we would argue that
there is also a strong SC fluctuation in LaRu$_3$Si$_2$. As for a
three dimensional system judged from our band structure
calculations, this kind of strong SC fluctuation may suggest that
the superfluid density is low. Another interesting point shown in
Fig.4(a) is that the normal state specific heat (SH) coefficient C/T
shows a non-linear dependence on T$^2$ down to the lowest
temperature (0.38 K). This is clearly deviating from the prediction
of the Debye model. Taking the slope of $C/T$ vs. $T^2$ from the low
temperature data, we get the Debye temperature $T_D$ = 284 K. The
phonon contribution calculated based on the Debye model $C_{Debye}
\propto (T/T_D)^3\int_0^{T_D/T}[e^4e^x/(e^x-1)^2]dx$ is shown by the
red dashed line. One can see that the Debye model is seriously
violated. It is naturally questioned whether this violation is
induced by some electron correlation effect. For a non-Fermi liquid
with three dimensionality, the enhanced electron-electron
interaction will give an extra contribution to the electronic
specific heat\cite{QCP} $C_{e-e}=-AT^nlnT$ with n = 1 to 3. This
gives the correction to the Fermi liquid description, $n$ = 1
corresponds to the case of strong correlation, like in Heavy fermion
systems\cite{QCP}, while n = 3 corresponds to a weak correlation.
Thus we fit the data with $C=\gamma_nT+\beta T^3-AT^nlnT$ and found
a very good fitting when n takes 2 and $\gamma_n=36.38 mJ/mol K^2$,
as shown by the solid line, leading to $\beta=1.416mJ/mol K^4$ and
$A=3.61 mJ/mol K^3$. Therefore we intend to conclude that the
electron correlations may play an important role in the system. In
Fig.4(b), we derived the electronic specific heat by subtracting the
normal state background measured at 5 T. One can see that the low-T
part of $C_e/T$ exhibits a  flat feature, indicating a full SC gap.
There is a small upturn of $C_e/T$ in the low-T region when the
field is very weak, see the data of 0.25 T and 0.5 T, which is
attributed to the Schotkky anomaly due to the paramagnetic centers.
Using an integral based on the BCS formula for electronic specific
heat, we fit the data at $\mu_0$H = 0, 0.25 and 0.5 T and show with
the solid lines. In this way we obtained the data in the zero
temperature limit for each field. For higher magnetic fields, it is
known that the Schottky anomaly becomes weaker, we can determine the
low-T data directly from the experiment data. Fig.4(c) shows the
temperature dependence of the entropy calculated using $S= \int_0^T
C_e/T'dT'$, it is clear that the entropy is conserved at T$_c$ as
judged by $S_S-S_N|_{T_c}$ = 0, where $S_S$ or $S_N$ are the
entropies of the superconducting state and the normal state
integrated up to T, and $S_S-S_N= \int _0^T (C_S/T'-C_N/T')dT'$.
After obtaining the low-T data, we derived the electronic SH at
different magnetic fields and plot them in Fig.4(d). Interestingly,
a non-linear field dependence can be easily seen here. Further
analysis finds that this non-linear dependence is actually different
from the $\sqrt{H}$ relation (shown by the solid line in Fig.4(d))
predicted for a clean superconductor with line-node gap, the data
below about 0.5 T seems to be more linear. Our result here is
certainly different from a linear relation as predicted for a single
isotropic SC gap. A multigap feature would already explain the data,
but as seen from the flattening of $C_e/T$ at T $\rightarrow$ 0, we
would argue that it is not the multigap feature, but the gap
anisotropy that leads to the non-linear field dependence of $C_e/T$
. A momentum resolved measurement is highly desired to uncover this
puzzle.

\begin{figure}
\includegraphics[width=8cm]{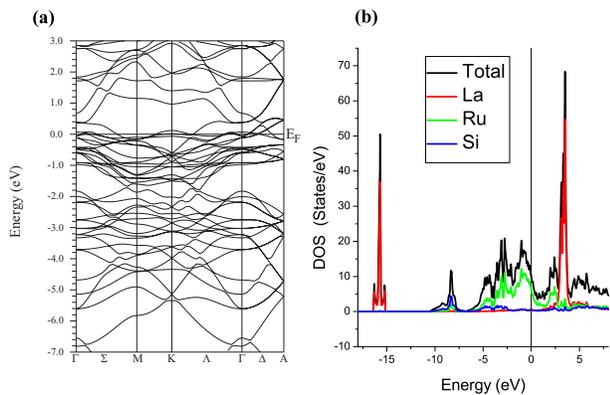}
\caption {(color online) (a) The energy bands obtained from the
DFT calculations. The dense bands near E$_F$ are derived from the
Ru 4d orbitals. (b) The electronic density of states (DOS) from the band
structure calculations. It is found that the DOS at the E$_F$ are
mainly contributed by the Ru orbitals. The DOS from the La and the
Si atoms at E$_F$ are negligible.} \label{fig5}
\end{figure}

In order to have a comprehensive understanding to the data, we did
the density-functional theory calculations by using the WIEN2k
package\cite{WIEN2k} utilizing the generalized gradient
approximation\cite{PBE} for the exchange-correlation potential. As
shown in Fig.5(a), the bands around Fermi level are mainly
contributed by Ru 4d. The Si 3p bands are very wide and have some
hybridization with Ru 4d. Further analysis of the calculation
shows that the crystal-field splitting upon Ru 4d orbitals is
quite weak, consequently all Ru 4d electron should play an important role in
the electron conduction and related superconductivity. There are several bands
crossing the Fermi level, which leads to complicated 3D Fermi
surfaces, this will be presented elsewhere. Since the band closed
to Fermi level is narrow, and the density of state at Fermi level
is high as shown in Fig.5 (b). We also perform spin polarized
calculations to check the possible magnetic instability. The
calculation shows that the ferromagnetism is not stable for this
compound. We cannot find any strong nesting effect in the
Fermi-surface, thus the SDW order is unlikely. Worthy of noting is
that all the five Ru 4d orbitals contribute to the conduction in
$LaRu_3Si_2$, which is very similar to the case of the iron in the
iron-pnictide superconductors\cite{Singh}. Actually a Ru-based
compound, namely $LaRu_2P_2$, is a superconductor with $T_c$ = 4.1
K, which has the similar structure of the $BaFe_{2-x}Co_xAs_2$
superconductor\cite{Mandrus,Canfield}, and probably they share the same
superconducting mechanism. This reminds us that the correlation
effect may play some roles in the superconductivity of
$LaRu_3Si_2$.

In summary, resistivity, magnetization and specific heat have been
measured in a Ru-based superconductor $LaRu_3Si_2$ with $T_c$ of
about 7.8 K. The temperature dependence of the normal state specific heat
coefficient C/T deviates clearly from the Debye model, but shows
the possible evidence of electron correlations. The
superconducting transitions measured by both resistivity and
specific heat self-consistently present the evidence of strong
superconducting fluctuations, resembling that in the cuprates. The
field induced quasiparticle density of states show a non-linear
magnetic field dependence, which is argued as a gap anisotropy.
Combining the novelties found both in the normal state and the
superconducting state, we argue that the electron correlations may
play an important role in the occurrence of superconductivity in
$LaRu_3Si_2$.

\begin{acknowledgments}
We appreciate the useful discussions with Jan Zaanen, Zidan Wang,
Zlatko Tesanovic, Tao Xiang, Qianghua Wang, and Jianxin Li. This
work is supported by the NSF of China (11034011), the Ministry of Science and
Technology of China (973 projects: 2011CBA001002, 2012CB821403, 2010CB923000).

\end{acknowledgments}

$^{\star}$ hhwen@nju.edu.cn

\end{document}